# Maskless off-axis X-ray Holography

Kahraman Keskinbora,[1,2,3,*] Abraham Levitan,[2] Gisela Schütz,[1] and Riccardo Comin[2,*]

[1] Max Planck Institute for Intelligent Systems, Stuttgart, Germany

[2] Massachusetts Institute of Technology, Cambridge, MA, USA

[3] Harvard University, John A. Paulson School of Applied Sciences, Center for Integrated Quantum Materials, Cambridge, MA, USA

*Corresponding authors: kahraman@mit.edu, rcomin@mit.edu

**Abstract**

Off-axis X-ray holography is a lensless imaging technique that allows unambiguous retrieval of an object's exit-wave function with high fidelity. It has been used with great success to study nanoscale phenomena and spatio-temporal dynamics in solids, with sensitivity to the phase component of electronic and magnetic textures. However, the method requires patterning a nanostructured holography mask directly onto the sample. This invasive fabrication process is labor-intensive and defines a fixed field of view, limiting the range of applicable samples and diminishing the signal-to-noise ratio at high-resolution. In this work, we propose using wavefront-shaping X-ray diffractive optics to create a spatially structured probe with full control of its phase at the sample plane, obviating the need for a holography mask. We demonstrate in silico that the method can image nanoscale structures and magnetic textures. The proposed holography route can image extended regions of interest, enabling investigation of a broad range of physical phenomena at the nanoscale including magnetism and electronic phase coexistence in quantum materials. We further envisage applications in phase contrast imaging of other weakly absorbing objects in the realm of soft and biological matter research.

1. **Introduction**

The improvements in phase-contrast imaging methods for applications ranging from materials science to medicine are coincident with the development of high-coherence and high-repetition-rate photon sources [1-3]. Thanks to the increasing interest from a multitude of key research fields, coherent imaging techniques are becoming increasingly popular. These phase-sensitive imaging methods are well-positioned to reap the scientific benefits of technical advances in high-coherence X-ray sources such as diffraction-limited synchrotrons and free-electron lasers [4] for ultra-high time resolution imaging at the nano- and mesoscale.

The phase of the exit-wave leaving a coherently illuminated sample contains critical information relating to the internal structure of the sample. However, this information is lost during detection. This so-called missing-phase problem is ubiquitous in optical, electron, and

X-ray microscopy. It has motivated a rich variety of imaging techniques using both X-rays and electrons that strive to recover the complex sample exit wave from recorded far-field intensity patterns. These methods include in-line [5], off-axis [6], and hybrid holography schemes [7], as well as coherent diffractive imaging (CDI) and ptychography approaches [8-15] based on iterative reconstruction algorithms [16, 17].

X-ray Fourier transform holography (FTH) is one of the most successful of these approaches in X-ray phase-contrast imaging [18, 19]. In the FTH approach, a holography mask composed of a micron-sized sample aperture and a nanometer-sized reference aperture is fabricated directly onto the sample. The major advantage of FTH is the simplicity of the reconstruction method, implemented using a single Fourier transform. It is also compatible with single-shot imaging methods. However, this comes at the steep price of a fixed, preselected sample region of interest, inefficient utilization of the total incident photon flux, and labor-intensive sample preparation. Several approaches have been proposed to improve some of these aspects of FTH [19-23]. For instance, to extend the field of view beyond a few microns given by the prefabricated sample window, Guehrs *et al.* separated the holography mask from the sample to image extended objects. Hessing *et al.* were able to overcome the field of view limitations by using a separate mask on a different stage in close proximity of the sample[21]. Geilhufe *et al.* utilized a Fresnel zone plate (FZP) instead of a reference aperture to overcome the diminishing amounts of light transmission from a nano-sized reference hole [22] and enhance the signal-to-noise ratio. However, while the latter improves the signal rate, the region of interest remains fixed. In an earlier implementation of X-ray FTH, McNulty *et al.* used the $0^{th}$ order light transmitting through an FZP to illuminate the sample while the focus of the FZP provided a reference point source [19]. More recently, Balyan and Haroutunyan devised an FTH scheme based on a two- or three-FZP interferometer to project the object and reference beams onto a sample, again with a fixed field of view and with reduced photon efficiency due to multiplicative impact of the zone plates' transmission rates [23]. Also recently, a beam shaping approach was proposed by Marchesini and Sakdinawat [24]. Their approach depends on modifications to the amplitude and phase at the plane of binary X-ray optic by varying the duty cycle, zone positions, harmonics, *etc*. to create the desired structured illumination at the focal plane. Among other imaging methods, holography is mentioned briefly as a possible application of these optics.

Here, we lay out and describe in detail all the necessary steps for removing dependence of off-axis X-ray holography on the hard-encoded holography mask. We combine different elements from digital holography, holographic image projection and holographic microscopic imaging to present an imaging scheme which utilizes structured illumination instead of a mask. All the aspects of this method, from the synthesis of the computer generated holograms to the simulation of the full FTH imaging cycle are explicitly defined, clarified, and studied.

In the approach we propose here, a beam-shaping optic, synthesized using a double-constraint Gerhcberg-Saxton algorithm [25], concentrates incident X-rays to form a structured illumination pattern at the sample plane. This pattern is composed of a large, uniformly illuminated sample beam and a tightly focused reference beam with a well-defined phase relationship between the two. By making use of this optic, we simultaneously address all three abovementioned challenges of the FTH method. A schematic summary of the full imaging scheme that we call **St**ructured **I**llumination **X**-ray **H**olography, **StIXH,** is shown in Figure 1a. The decoupling between the holographic (virtual) mask and the sample opens new experimental possibilities that are otherwise impossible, as will be discussed later.

## 2. Synthesis of the Computer Generated Holograms

Holographic image projection is an active field of research with applications in 3D displays [26] and acoustic medical imaging [27] and X-ray beam shaping [24]. In many holographic image projection applications, computer-generated holograms only project the designed intensity pattern, without tailoring the phase at the projection plane. The projected intensity profiles usually suffer from the so-called speckle-noise, caused by random phase distribution within the field. This speckle-noise is detrimental if a structured illumination is to be used for the holographic microscopic imaging. Numerous speckle-noise reduction methods are available in holographic projection, broadly falling into two categories: optical and numeric [28]. Optical routes for speckle suppression depend on reducing the coherence, for instance, by using diffusers [29]. However, to achieve X-ray holography using structured light, we simultaneously need tailored phase and amplitude profiles (so-called complex-amplitude control) and a high degree of coherence at the sample plane. Therefore, the above-mentioned coherence reduction approaches cannot be applied here.

The possibility of simultaneously tailoring an amplitude and phase field has been explored in the context of the digital image projection to reduce the speckle-noise. Such approaches include using metasurfaces, spatial light modulators (SLMs) or using a modified Gerchberg-Saxton Algorithm [25, 30, 31] to control not only the amplitude but also the phase of the projected beam. This approach is called complex-amplitude control.

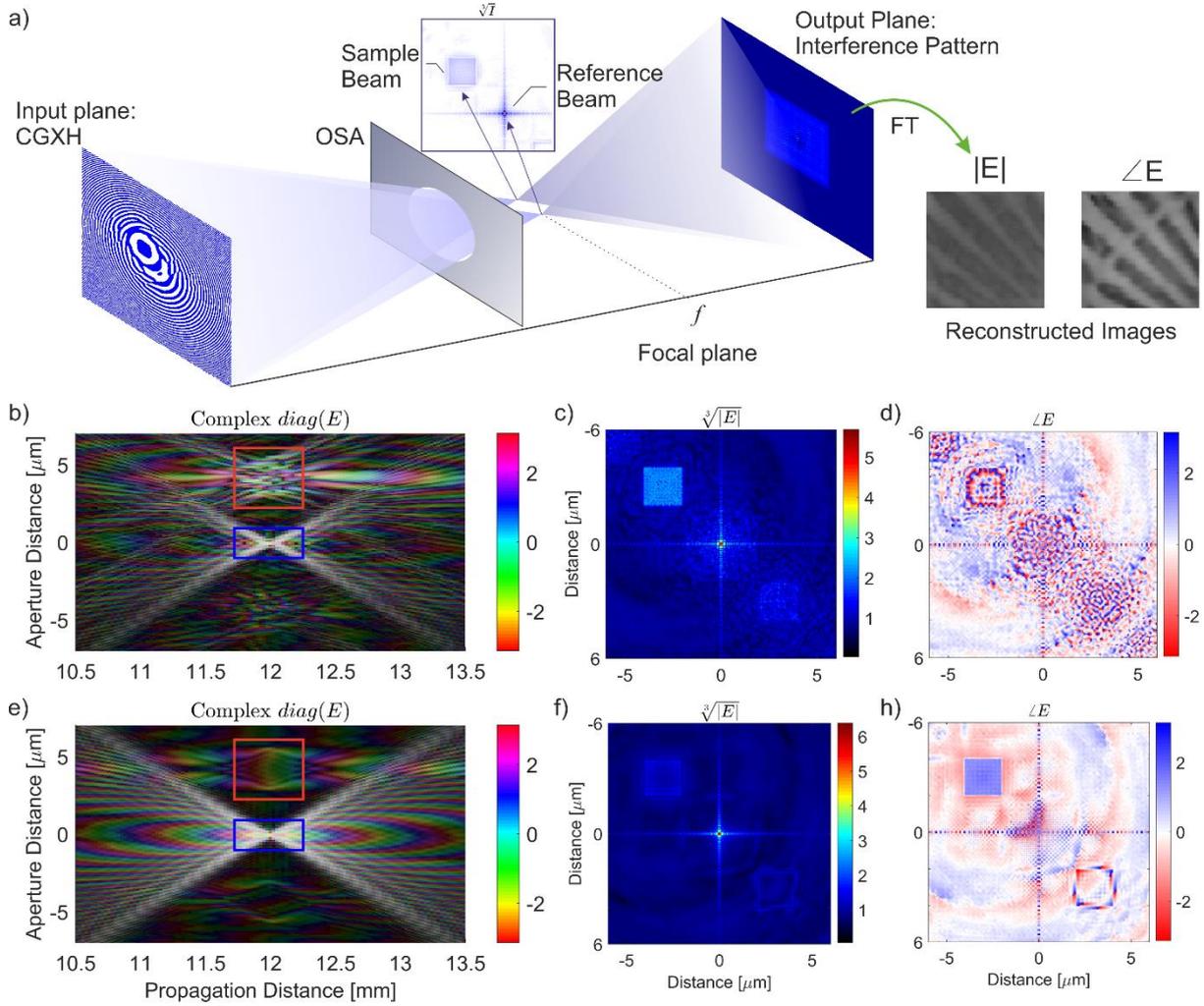

**Figure 1 | Schematic description of the proposed experiment. a** In the StIXH experiment, a computer-generated X-ray hologram is illuminated by a coherent plane wave. Two separate beams with tailored complex-amplitudes are simultaneously projected onto the sample in lieu of the transmission mask, by a computer-generated hologram. A spatial filter called an order sorting aperture eliminates the spurious diffraction orders. The sample is illuminated by structured light at the focal plane (*f*), and the sample beam is imprinted with sample information such as chemical, structural, electronic/magnetic properties. An area detector captures the far-field interference pattern of the sample and the reference beams. A Fourier transform of the interference pattern yields the amplitude and phase images of the sample. **b** A 2D diagonal slice of the complex field created by a hologram synthesized using GSA, along the propagation direction near the focal plane (12 mm). Note the scrambled phase of the sample beam and the intensity variation (marked red) in relation to the reference beam (marked blue). **c, d** The amplitude and phase of the structured illumination pattern of a GSA hologram at *f* exhibit speckle noise in the sample beam as a result of constructive and destructive interference within this region due to the random phase. **e** Slice of the complex field of a DC-GSA hologram at *f*. Notice how the sample-beam has a flat phase at the focal plane in contrast to **b. f, h** The amplitude and phase of DC-GSA holograms at *f* clearly show relatively flat phase and intensity distribution.

At X-ray wavelengths, spatial light modulators and meta-surfaces [30, 31] are not practical for speckle suppression using complex amplitude control. Hence, we follow a modified double-constraint [25] Gerchberg-Saxton algorithm (DC-GSA) [32] approach for synthesizing the computer-generated X-ray holograms (CGH) as binary diffractive optics. We synthesize the CGH by defining a *signal domain* composed of the sample and the reference beams in the probe field. Here, we tailor both the amplitude and the phase of the beam. The rest of the probe field is composed of the *freedom domain* where the amplitude is ideally 0, and the phase can take any value. During each iteration cycle of DC-GSA, the amplitude and phase of the probe field are updated within the signal domain in an attempt to match the target amplitude and phase. In contrast, within the freedom the domain, the probe's amplitude is dampened by a predefined factor without altering the phase. In the optic plane, a square mask is forced that defines the optic boundaries. The CGH that projects the desired probe field in the sample plane takes form after around 20 iterations (see Figure S2. For a detailed explanation of the CGH synthesis scheme, see the SI).

The resulting X-ray CGH allows control over the intensity and the phase of the light projected onto the focal plane. When the CGH is illuminated with a plane wave, it forms a structured illumination pattern at the focal plane. An instance of such an experiment is depicted in Fig. 1a. The differences in the projected light-field caustics between a conventional GSA and modified DC-GSA algorithms are presented in Figs. 1b through h. The complex field near the focal plane along the propagation direction from CGH synthesized by conventional GSA exhibits a scrambled phase profile as shown in Fig. 1b. The sample beam has abrupt intensity variations (Fig. 1c) due to its random phase distribution (Fig. 1d). In comparison, the CGHs generated by DC-GSA controls both the amplitude and phase of the *P,* as shown by a diagonal slice of the propagated beam in Fig.1e. Both the amplitude (Fig. 1f) and the phase (Fig. 1h) at the focal plane are relatively flat and only exhibit low-frequency modulations.

We estimated the focusing efficiency of such a hologram made out of 280 nm thick gold layer to be about 10 % at 1200 eV, which is well within the requirements of a state-of-the art X-ray microscope (See SI). The total intensity contained within the reference and the sample beam regions accounts for more than 87 % of the total intensity behind the OSA in the focal plane *P*.

## 3. Imaging simulations

Two imaging case studies involving a nanostructured test object and magnetic worm domains in synthetic samples are used for an *in silico* demonstration of the StIXH's capabilities. The simulations are conducted using empirical charge and magnetic scattering factors data taken from the literature [33, 34].

*Structural Imaging* – We first consider imaging simulations of a Siemens star test object with nano-sized features. The synthetic sample is a binarized SEM micrograph (Fig. 2a), which was then assigned complex refractive index values from the CXRO database [33], and a thickness of 180 nm. The thin sample approximation was used to calculate the sample exit-wave, which (Fig. 2b) was then propagated 16 cm downstream to the detector plane using a scaled Fraunhofer propagator [35]. Only the intensity of the interference pattern, $I = |E(x,y)|^2$, was kept (Fig. 2c). The magnitude and phase of the sample exit-wave were then calculated by an FFT of the intensity and shown in Fig. 2d and Fig. 2e, respectively. Both the amplitude and the phase of the exit-wave were reconstructed with fidelity to the original object. The spokes in the second ring structure of the Siemens star are clearly displayed. It is worth noting here that the reconstruction is quite robust against detector shot noise. The reconstruction quality improves significantly above $10^1$ photons/pixel, and reliable reconstructions can be achieved starting from a $10^2$ maximum photons/pixel in the detector plane (Figure S3). For the following StIXH simulation, a maximum of $10^3$ photons/pixel was used.

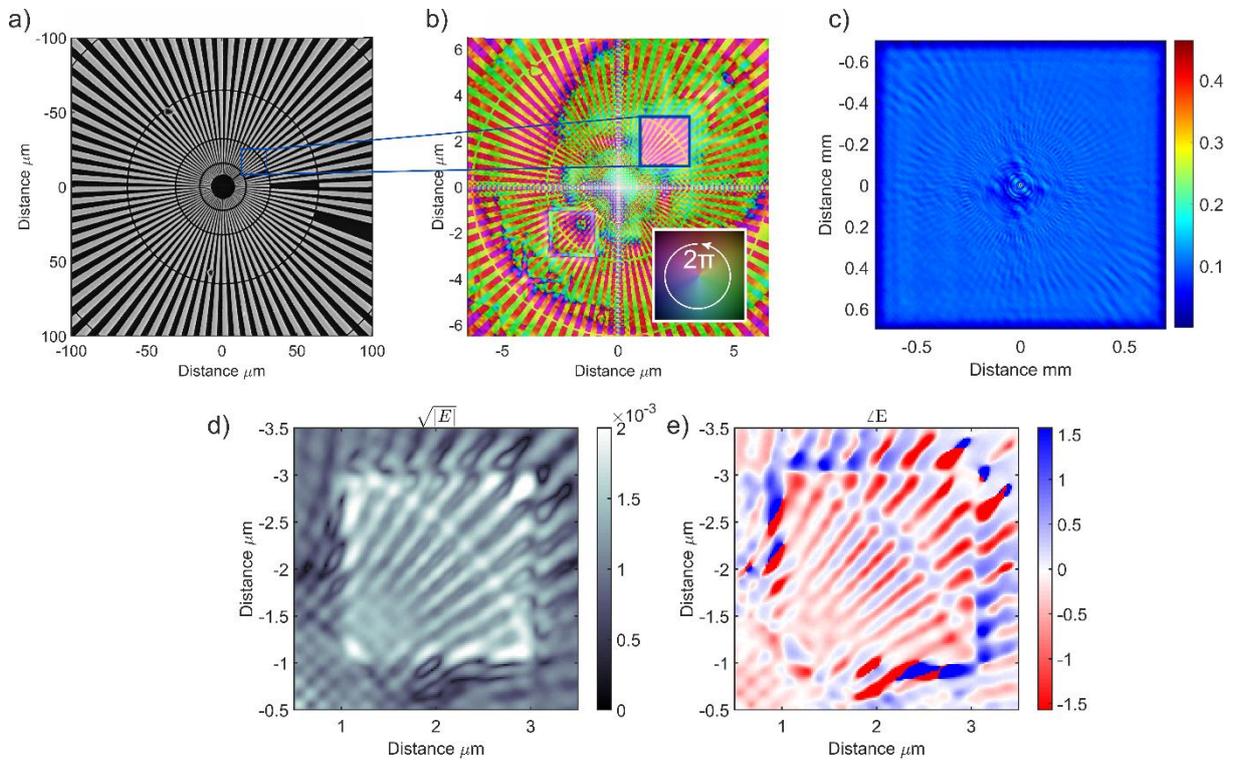

**Figure 2 | Simulated StIXH imaging using synthetic data. a** An SEM micrograph of the Siemens star that was used to prepare the synthetic data. The SEM image was binarized and downsampled to match the simulated support size. The overlay shows the position of the sample beam. The sample and lens gold thicknesses were taken as 180 nm and 280 nm, respectively. The refractive index data were taken from http://cxro.lbl.gov/ [33]. **b** The complex exit wave function shows the reference beam as the bright spot. The sample beam is imprinted with the sample information at this point. The inset shows the color-wheel, with brightness proportional to the intensity and hue to the phase. **c** The interference pattern as recorded 16 cm downstream from the sample plane, stretched with a cube root to exhibit its features more clearly. Only the intensity of the interference pattern is recorded at the detector plane, and the phase is lost. A scaled Fourier transform of the interference intensity distribution leads to the reconstructed amplitude (**d**) and phase (**e**) of the sample ROI illuminated by the sample-beam.

*Magnetic Imaging* –Next, we turn our attention to data-driven simulations of synthetic magnetic materials with worm-domains (For details see SI Fig. S4). For the case of out-of-plane magnetization in a metal alloy material probed in the forward scattering direction by circularly polarized X-rays, the scattering factor simplifies [36] to $f^n \approx f_c^n + if_m^n \begin{pmatrix} im_z^n & 0 \\ 0 & -im_z^n \end{pmatrix} \approx f_c^n \pm f_m^n m_z^n$ for LC-to-LC (+) and RC-to-RC (-) polarization channels, where $m_z$ is the magnetization vector, $f_c'$, $f_c''$ and $f_m'$, $f_m''$ are the real and complex parts of the charge and magnetic scattering factors, respectively. Finally, LC and RC stands for left and right circular polarization, respectively. For the simulations shown in Figure 3, complete out of plane magnetization is assumed and as a result the scalar diffraction theory provides a satisfactory approximation.

In Figure 3a, a synthetic magnetic sample with a totally out-of-plane $m_z$ is shown. The sample is modeled as a 50 nm thick CoFeB thin film by including measured $f_c$ and $f_m$ data (Figure 3b). For a full history of the processing of the synthetic magnetic sample, please see SI Figure S5. Image reconstruction results at three different energies, 706.9 eV where $f_m'$ is maximum, 707.7 eV where $f_m''$ is maximum, and $f_m'$ is almost zero, and at 708.7 eV with intermediate $f_m$ values, are given in Figure 3d, e, and f, respectively. As expected, phase contrast is strongest where $f_m'$ is the largest and disappears at 707.7 eV where $f_m'$ approaches 0. Note that the sample is uniform in composition and thickness; therefore, the $f_c$ does not contribute to the contrast but results in a uniform absorption and phase shift. Further, magnetic structure factors are markedly smaller than charge scattering factors (Figure 3b).

Despite significant detector shot noise, the phase and amplitude of the object exit-wave successfully capture the general features of the magnetic texture. A high spatial frequency modulation that comes from the probe itself is visible, particularly in the amplitude reconstructions. In contrast, a cleaner reconstruction of the phase was achieved. Removal of such probe-based artifacts from the image reconstruction *via* probe calibration by an initial ptychography scan before moving to the actual StIXH experiment or using an iterative reconstruction algorithm will be considered for future studies.

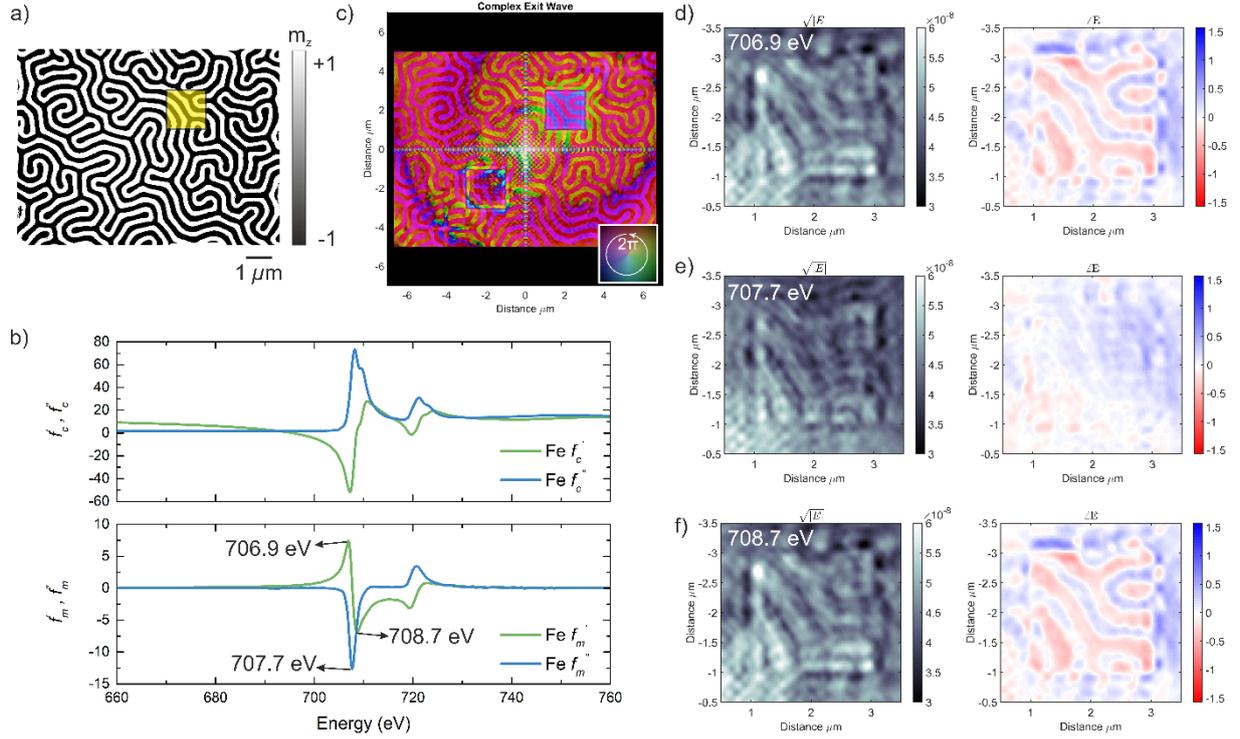

**Figure 3 | Imaging simulations on magnetic worm domains using StIXH. a** The domain wall structure represents the out-of-plane magnetization vectors (image source: [37]). The overlaid square shows the approximate region that is spanned by the sample beam. **b** The charge and magnetic structure factors for iron in CoFeB (Data courtesy of D. B. Boltje and E. Goering) [34]. **c** The complex exit wave function at the sample plane. Note that both the reference and the sample beam pass through the sample in this case. **d**, **e**, and **f**, The reconstructed square root of the amplitude and phase of the magnetic worm domain patterns at 706.9, 707.7, and 708.7 eV, respectively. CoFeB thickness was taken as 50 nm. Detector shot noise with a maximum of 1000 ph/pix.

## 4. Discussion

The conventional FTH method provides excellent stability through mechanical integration of the illumination mask and sample at the cost of: (i) limiting the sample type and field of view; (ii) inefficient usage of coherent flux; and (iii) a rather labor-intensive sample preparation.

The method proposed here addresses precisely these limitations of conventional X-ray holography and provides further experimental advantages: (i) the sample and the reference beams are not defined by pre-patterned structures; (ii) The field of view is not physically anchored to a preselected region of interest, allowing free investigation of extended samples; (iii) Thanks to this separation, the proposed method is suitable not only for transmission experiments but also for a reflection geometry, enabling Bragg holography at finite momentum transfers to visualize collective phases of matter with long-range order, including charge density waves, antiferromagnetic order, topological textures (skyrmions and polar vortices), and others that are peculiar to strongly interacting electron systems; (iv) StIXH is compatible with single-shot, ultrafast spatiotemporal imaging of non-periodic dynamics at the nanoscale, which is not possible via stroboscopic dynamic imaging; (v) In StIXH, the relative intensity at the sample and reference beams can be freely varied; (vi) StIXH uses coherent incident flux more efficiently compared to FTH, where the vast proportion of incoming photons are discarded.

Alongside its promises, StIXH comes with specific challenges. Most importantly, the incompatibility with a beamstop (See Fig. S5) is likely to be an important obstacle. As the quality of the reconstruction depends on the absence of residual light that illuminates undesired portions of the sample, finding an alternative approach to blocking stray light is critical. One possibility to overcome the issue is to increase the efficiency of the optic by resorting to kinoform type X-ray CGHs, with three-dimensional surface-relief structures, which can be fabricated via gray-scale ion beam lithography [38-41] or 3D nanoprinting [42]. Another approach could be to encode a secondary phase ramp into the CGH to separate the unwanted $0^{th}$ order, from the actual probe field [24, 43]. Then the $0^{th}$ order light can be blocked by an external beamstop.

The discussed CGH generation method can also be used for tailoring the probe wave-function with arbitrary complexity. In addition to its uses in X-ray holography, the same approach can be used to generate specialized probes such as vortex beams or for ptychography, coherent diffractive imaging, and other phase-contrast X-ray imaging methods. Furthermore, achieving full control over the phase as well as the amplitude of a structured illumination *via* a simple monolithic optic has a broad range of applications in acoustic and light-based medical imaging technologies [27] as well as untethered micro-robotic and micromanipulation applications [44-46].

## 5. Conclusions

We have applied the DC-GSA algorithm to the synthesis of efficient X-ray holograms that can project arbitrarily structured illumination with simultaneous phase and amplitude control. With proper design, the optic can concentrate the incident light into an illumination pattern that replaces the physical holography mask in conventional FTH experiments, addressing various limitations of FTH. Through *in silico* studies, it was shown that the proposed imaging modality can be used to visualize structural and magnetic textures, and is robust against detector shot noise. The increased intensity on the sample plane allows for efficient usage of the coherent flux and positions this method very well for reaping the scientific benefits of current and future high-coherence X-ray sources. The light concentration capability of the diffractive holograms used for StIXH is expected to be a particular advantage for less intense sources, such as high harmonic generation sources where the brightness decreases as the output photon energy increases into the soft X-ray range. Hence, holography experiments using HHG sources will markedly benefit from these proposed condenser optics.

Its single-shot reconstruction capability promises ultra-high time resolution measurements of non-reproducible dynamics using high repetition rate X-ray sources. Furthermore, by decoupling the illumination function from the object, microscopic investigations using X-ray holography will not be limited to preselected regions of interest but rather the whole, extended sample such as biological cells, or electronic and magnetic devices.

## 6. Acknowledgments

K.K. thanks Prof. Stefan Eisebitt of the Max Born Institute and Dr. Cigdem Ozsoy-Keskinbora of the Harvard University's CIQM for fruitful discussions on off-axis holography and Prof. Robert Westervelt for providing access to CIQM's infrastructure. Authors also thank Dr. Eberhard Goering and Mr. Daan B. Boltje of the Max Planck Institute for Intelligent Systems for kindly providing the experimental values of charge and magnetic structure factors for CoFeB. K.K. was supported by the German Research Foundation (DFG) under project number 428809035. Work at MIT (A.L. and R.C.) has been supported by the Department of Energy, Office of Science, Office of Basic Energy Sciences, under Award Number DE-SC0019126. K.K. also acknowledges partial support by the STC Center for Integrated Quantum Materials, NSF Grant No. DMR-1231319.

## 7. References

1. R. Gradl, M. Dierolf, B. Günther, L. Hehn, W. Möller, D. Kutschke, L. Yang, M. Donnelley, R. Murrie, A. Erl, T. Stoeger, B. Gleich, K. Achterhold, O. Schmid, F. Pfeiffer, and K. S. Morgan, "In vivo Dynamic Phase-Contrast X-ray Imaging using a Compact Light Source," Scientific Reports **8**, 6788 (2018).
2. Z. Najmudin, S. Kneip, M. S. Bloom, S. P. D. Mangles, O. Chekhlov, A. E. Dangor, A. Döpp, K. Ertel, S. J. Hawkes, J. Holloway, C. J. Hooker, J. Jiang, N. C. Lopes, H. Nakamura, P. A. Norreys, P. P. Rajeev, C. Russo, M. J. V. Streeter, D. R. Symes, and M. Wing, "Compact laser accelerators for X-ray phase-contrast imaging," Philosophical Transactions of the Royal Society A: Mathematical, Physical and Engineering Sciences **372**(2014).
3. M. P. Olbinado, X. Just, J.-L. Gelet, P. Lhuissier, M. Scheel, P. Vagovic, T. Sato, R. Graceffa, J. Schulz, A. Mancuso, J. Morse, and A. Rack, "MHz frame rate hard X-ray phase-contrast imaging using synchrotron radiation," Opt. Express **25**, 13857-13871 (2017).
4. M. Yabashi and H. Tanaka, "The next ten years of X-ray science," Nat Photon **11**, 12-14 (2017).
5. D. Gabor, "A new microscopic principle,"  (Nature Publishing Group, 1948).
6. G. Möllenstedt and H. Düker, "Beobachtungen und Messungen an Biprisma-Interferenzen mit Elektronenwellen," Zeitschrift für Physik **145**, 377-397 (1956).
7. C. Ozsoy-Keskinbora, C. B. Boothroyd, R. E. Dunin-Borkowski, P. A. van Aken, and C. T. Koch, "Hybridization approach to in-line and off-axis (electron) holography for superior resolution and phase sensitivity," Scientific Reports **4**, 7020 (2014).
8. F. Pfeiffer, "X-ray ptychography," Nat. Photonics **12**, 9-17 (2018).

9. C. Donnelly, V. Scagnoli, M. Guizar-Sicairos, M. Holler, F. Wilhelm, F. Guillou, A. Rogalev, C. Detlefs, A. Menzel, J. Raabe, and L. J. Heyderman, "High-resolution hard x-ray magnetic imaging with dichroic ptychography," Phys. Rev. B **94**, 064421 (2016).
10. P. Thibault, M. Dierolf, O. Bunk, A. Menzel, and F. Pfeiffer, "Probe retrieval in ptychographic coherent diffractive imaging," Ultramicroscopy **109**, 338-343 (2009).
11. J. M. Rodenburg, "Ptychography and Related Diffractive Imaging Methods," in *Advances in Imaging and Electron Physics*, Hawkes, ed. (Elsevier, 2008), pp. 87-184.
12. D. A. Shapiro, Y.-S. Yu, T. Tyliszczak, J. Cabana, R. Celestre, W. Chao, K. Kaznatcheev, A. L. D. Kilcoyne, F. Maia, S. Marchesini, Y. S. Meng, T. Warwick, L. L. Yang, and H. A. Padmore, "Chemical composition mapping with nanometre resolution by soft X-ray microscopy," Nat. Photonics **8**, 765-769 (2014).
13. X. Shi, N. Burdet, B. Chen, G. Xiong, R. Streubel, R. Harder, and I. K. Robinson, "X-ray ptychography on low-dimensional hard-condensed matter materials," Applied Physics Reviews **6**, 011306 (2019).
14. Z. Chen, M. Odstrcil, Y. Jiang, Y. Han, M.-H. Chiu, L.-J. Li, and D. A. Muller, "Mixed-state electron ptychography enables sub-angstrom resolution imaging with picometer precision at low dose," Nat. Commun. **11**, 2994 (2020).
15. Y. Jiang, Z. Chen, Y. Han, P. Deb, H. Gao, S. Xie, P. Purohit, M. W. Tate, J. Park, S. M. Gruner, V. Elser, and D. A. Muller, "Electron ptychography of 2D materials to deep sub-ångström resolution," Nature **559**, 343-349 (2018).
16. J. R. Fienup, "Phase retrieval algorithms: a personal tour [Invited]," Appl. Opt. **52**, 45-56 (2013).
17. J. R. Fienup, "Reconstruction of an object from the modulus of its Fourier transform," Opt. Lett. **3**, 27-29 (1978).
18. S. Eisebitt, J. Luning, W. F. Schlotter, M. Lorgen, O. Hellwig, W. Eberhardt, and J. Stohr, "Lensless imaging of magnetic nanostructures by X-ray spectro-holography," Nature **432**, 885-888 (2004).
19. I. McNulty, J. Kirz, C. Jacobsen, E. H. Anderson, M. R. Howells, and D. P. Kern, "High-Resolution Imaging by Fourier Transform X-ray Holography," Science **256**, 1009-1012 (1992).
20. E. Guehrs, C. M. Güunther, B. Pfau, T. Rander, S. Schaffert, W. F. Schlotter, and S. Eisebitt, "Wavefield back-propagation in high-resolution X-ray holography with a movable field of view," Opt. Express **18**, 18922-18931 (2010).
21. P. Hessing, B. Pfau, E. Guehrs, M. Schneider, L. Shemilt, J. Geilhufe, and S. Eisebitt, "Holography-guided ptychography with soft X-rays," Opt. Express **24**, 1840-1851 (2016).
22. J. Geilhufe, B. Pfau, M. Schneider, F. Buttner, C. M. Gunther, S. Werner, S. Schaffert, E. Guehrs, S. Frommel, M. Klaui, and S. Eisebitt, "Monolithic focused reference beam X-ray holography," Nat Commun **5**(2014).
23. M. Balyan and L. Haroutunyan, "X-ray Fourier transform holography by amplitude-division-type Fresnel zone plate interferometer," J. Synchrotron. Radiat. **25**, 241-247 (2018).
24. S. Marchesini and A. Sakdinawat, "Shaping coherent x-rays with binary optics," Opt. Express **27**, 907-917 (2019).
25. C. Chang, J. Xia, L. Yang, W. Lei, Z. Yang, and J. Chen, "Speckle-suppressed phase-only holographic three-dimensional display based on double-constraint Gerchberg-Saxton algorithm," Appl. Opt. **54**, 6994-7001 (2015).
26. G. Makey, Ö. Yavuz, D. K. Kesim, A. Turnalı, P. Elahi, S. Ilday, O. Tokel, and F. Ö. Ilday, "Breaking crosstalk limits to dynamic holography using orthogonality of high-dimensional random vectors," Nat. Photonics **13**, 251-256 (2019).
27. S. Jiménez-Gambín, N. Jiménez, J. M. Benlloch, and F. Camarena, "Holograms to Focus Arbitrary Ultrasonic Fields through the Skull," Physical Review Applied **12**, 014016 (2019).
28. V. Bianco, P. Memmolo, M. Leo, S. Montresor, C. Distante, M. Paturzo, P. Picart, B. Javidi, and P. Ferraro, "Strategies for reducing speckle noise in digital holography," Light: Science & Applications **7**, 48 (2018).
29. F. Wyrowski and O. Bryngdahl, "Speckle-free reconstruction in digital holography," Journal of the Optical Society of America A **6**, 1171-1174 (1989).
30. G.-Y. Lee, G. Yoon, S.-Y. Lee, H. Yun, J. Cho, K. Lee, H. Kim, J. Rho, and B. Lee, "Complete amplitude and phase control of light using broadband holographic metasurfaces," Nanoscale **10**, 4237-4245 (2018).

31. E. Ulusoy, L. Onural, and H. M. Ozaktas, "Full-complex amplitude modulation with binary spatial light modulators," Journal of the Optical Society of America A **28**, 2310-2321 (2011).
32. R. W. Gerchberg and W. O. Saxton, "A Practical Algorithm for the Determination of Phase from Image and Diffraction Plane Pictures," Optik **35**, 237-246 (1972).
33. B. L. Henke, E. M. Gullikson, and J. C. Davis, "X-Ray Interactions: Photoabsorption, Scattering, Transmission, and Reflection at E = 50-30,000 eV, Z = 1-92," At. Data Nucl. Data Tables **54**, 181-342 (1993).
34. D. B. Boltje, *Voltage Induced Near Interface Changes of the Magnetocrystalline Anisotropy Energy: A study by X-ray Resonant Techniques in Combination with Conventional Magnetometry* (Universität Stuttgart, 2017).
35. D. G. Voelz, *Computational fourier optics: a MATLAB tutorial* (SPIE, 2011).
36. C. Donnelly, V. Scagnoli, M. Guizar-Sicairos, M. Holler, F. Wilhelm, F. Guillou, A. Rogalev, C. Detlefs, A. Menzel, and J. Raabe, "High-resolution hard x-ray magnetic imaging with dichroic ptychography," Phys. Rev. B **94**, 064421 (2016).
37. D. M. Gualtieri, "Magnetic_stripe_domains.jpg" (2011), retrieved 10.02.20, https://commons.wikimedia.org/wiki/File:Magnetic_stripe_domains.jpg.
38. K. Keskinbora, C. Grévent, M. Hirscher, M. Weigand, and G. Schütz, "Single-Step 3D Nanofabrication of Kinoform Optics via Gray-Scale Focused Ion Beam Lithography for Efficient X-Ray Focusing," Adv. Opt. Mater. **3**, 792-800 (2015).
39. K. Keskinbora, U. T. Sanli, C. Grévent, and G. Schütz, "Fabrication and x-ray testing of true kinoform lenses with high efficiencies,"  (Proc. SPIE, 2015), pp. 95920H-95920H-95926.
40. U. T. Sanli, H. Ceylan, I. Bykova, M. Weigand, M. Sitti, G. Schütz, and K. Keskinbora, "3D Nanoprinted Plastic Kinoform X-Ray Optics," Adv. Mater. (Weinheim, Ger.) **30**, 1802503 (2018).
41. I. Mohacsi, P. Karvinen, I. Vartiainen, A. Diaz, A. Somogyi, C. M. Kewish, P. Mercere, and C. David, "High efficiency x-ray nanofocusing by the blazed stacking of binary zone plates," in Proc. SPIE 2013), 88510Z-88510Z-88518.
42. U. T. Sanli, C. Jiao, M. Baluktsian, C. Grévent, K. Hahn, Y. Wang, V. Srot, G. Richter, I. Bykova, and M. Weigand, "3D Nanofabrication of High-Resolution Multilayer Fresnel Zone Plates," Advanced Science, 1800346 (2018).
43. H. Wang, Y. Liu, Q. Ruan, H. Liu, R. J. H. Ng, Y. S. Tan, H. Wang, Y. Li, C.-W. Qiu, and J. K. W. Yang, "Off-Axis Holography with Uniform Illumination via 3D Printed Diffractive Optical Elements," Advanced Optical Materials **7**, 1900068 (2019).
44. A. Aghakhani, O. Yasa, P. Wrede, and M. Sitti, "Acoustically powered surface-slipping mobile microrobots," Proceedings of the National Academy of Sciences **117**, 3469-3477 (2020).
45. H. Shahsavan, A. Aghakhani, H. Zeng, Y. Guo, Z. S. Davidson, A. Priimagi, and M. Sitti, "Bioinspired underwater locomotion of light-driven liquid crystal gels," Proceedings of the National Academy of Sciences **117**, 5125-5133 (2020).
46. K. Melde, A. G. Mark, T. Qiu, and P. Fischer, "Holograms for acoustics," Nature **537**, 518-522 (2016).

# Supporting Information

# Maskless off-axis X-ray holography

Kahraman Keskinbora,[1,2,3,*] Abraham Levitan,[2] Gisela Schütz,[1] and Riccardo Comin[2,*]

[1] Max Planck Institute for Intelligent Systems, Stuttgart, Germany

[2] Massachusetts Institute of Technology, Cambridge, MA, USA

[3] Harvard University, John A. Paulson School of Applied Sciences, Center for Integrated Quantum Materials, Cambridge, MA, USA

*Corresponding authors: keskinbora@is.mpg.de, rcomin@mit.edu

**Explanation of the reconstruction and simulation framework**

*Model* – Three planes are of interest in the modeling of the proposed experiment, as shown schematically in Fig. 1a. The input plane (i) contains the CGH, $\Psi_i(x,y)$, which was calculated using an iterative double-constraint algorithm from a target focal plane structured light distribution with tailored amplitude and phase. The focal plane, $f$, (ii) is where the structured probe, $P(x,y)$, interacts with the sample object function, $O(x,y)$. Finally, the detector plane (iii) intensity, $I(x,y)$ captures the interference pattern. The field $P(x,y)$ was simulated by propagating the field from the CGH input plane using the Fourier Beam Propagation Method [1, 2]. Scalar Rayleigh-Sommerfeld (RS) propagation was used in the iterative synthesis of the CGH input plane (i), which is initialized as a random phase distribution within a window that defines the aperture. At the focal plane (ii), two domains are established [3]. One is the signal domain, and the other is the so-called freedom domain. In the signal domain, both the amplitude and the phase of the tailored illumination pattern are forced. In the freedom domain, the amplitude is dampened by a factor at each iteration while the phase is free to take any value. Apart from ringing artifacts, the resultant phase is mostly random in this freedom domain, and the amplitude approaches 0. In the signal domain, both the amplitude and the phase approach the designed pattern. The two domains are then combined and backpropagated to the input plane (i). At each iteration step only the phase of the CGH, $(\angle\Psi_{CGXH})$, was kept and a support constraint that defines the width of the optic is forced in the input plane, defining a hologram with a clear aperture. After 20 to a 100 iteration cycles the resulting CGH phase was binarized with a 0.5-duty-cycle, assigned values such as thickness (280 nm) and actual material (Au) parameters. The optic exit-wave was modeled by considering a plane wave illumination under thin-sample conditions and was propagated to the focal plane, $f$, using the RS-function to calculate $P$, first by propagating to the OSA plane, and multiplying with the OSA aperture function and then propagated to the focal plane. Again, by using thin-sample approximation, the sample exit-wave was modeled as a Hadamard product of the probe and the object functions, $P(x, y) \bullet O(x, y)$. Finally, the holography experiment was simulated by propagating the exit-wave from the sample located in the focal plane (ii) to the detector plane (iii) *via* a scaled Fraunhofer (far-field) approximation propagator [1], which is valid as the far-field condition is satisfied (Fresnel number, $N_F \ll 1$). The intensity of the propagated wavefield at the detector plane is then recorded as $I(x, y)$, The reconstruction of the exit-wave is identical to the conventional FTH: an FFT of the $I(x, y)$ yields the autocorrelation of the sample exit-wave from itself. From this autocorrelation, the cross-correlation between the sample and reference beams can be extracted as an approximation to the real and imaginary parts of the sample exit-wave. (Fig. S1). The CGHs shown in Figure 1 and used in subsequent simulations were synthesized with 20 iteration cycles, as further iteration steps did not improve the root-mean-square error (Figure S2). The CGH's structure is compatible with conventional lithography methods [4-7]. It was found that the *P* is highly sensitive to the inclusion of a beamstop (Figure S3). The following imaging simulations do not include a beamstop at the optic plane. However, strategies to alleviate the challenges that come with the lack of a beamstop are discussed later on.

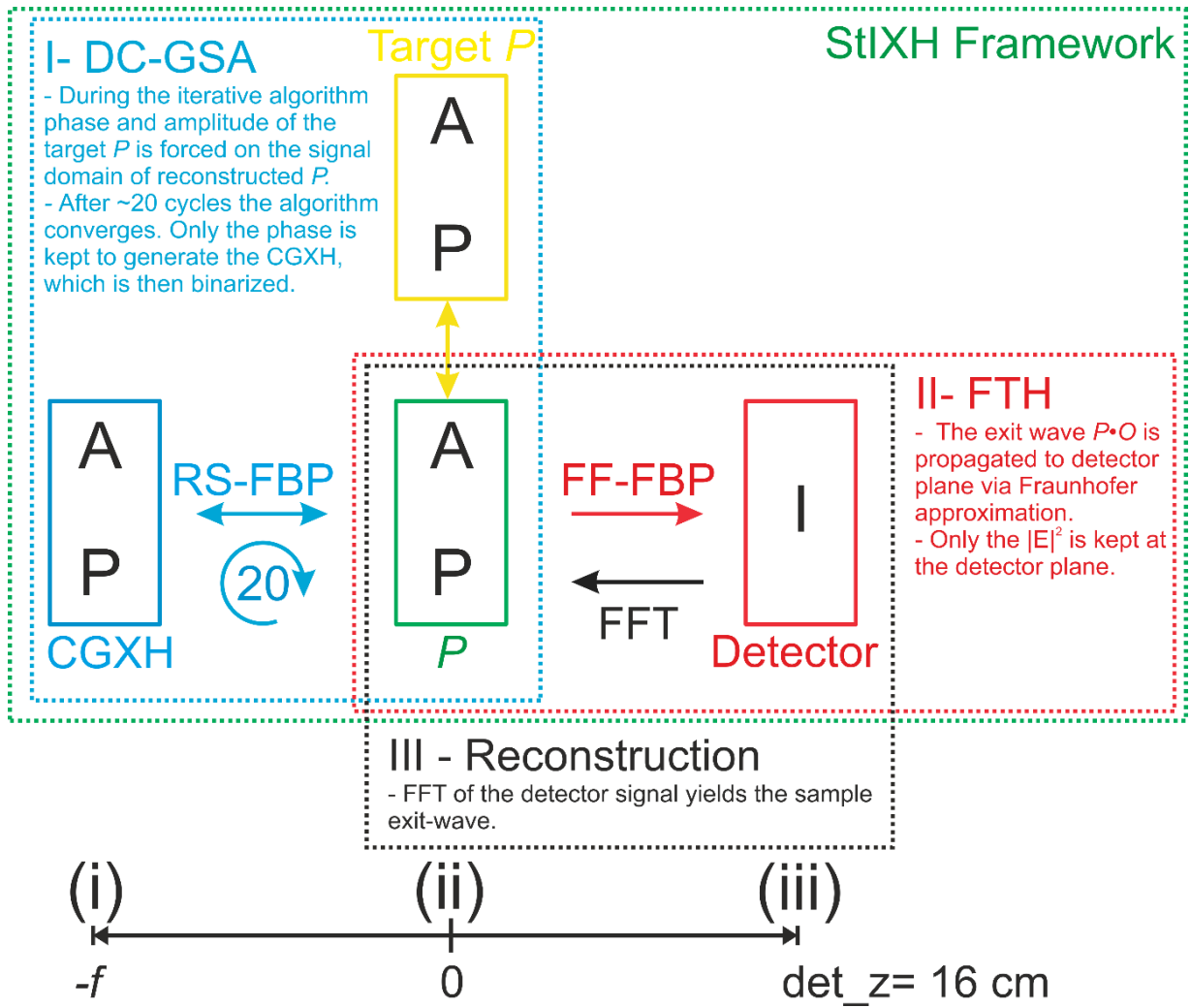

**Figure S1 | Overview of the simulation framework used in the study.** The region I represents the DC-GSA algorithm. The target probes amplitude and phase is forced in domain *P* only in regions we are interested in controlling the illumination properties, as described by Chang *et al*. [3]. We used scalar Rayleigh-Sommerfeld FBPM for this region. The region II represents the forward scattering experiment from sample exit-wave plane to the detector plane. Here, the Fraunhofer approximation is valid. The phase is discarded at the detector plane by considering the intensity, $I = |E(x,y)|^2$. The region III represents the reconstruction process, $P \bullet O = FFT(I(x,y))$, which is essentially a back-propagation of the interference intensity to the sample exit-wave plane.

**Pseudo-code for the Double Constraint Gerchberg-Saxton Algorithm**

```
for i = 1:1:iterations

    A = support.*exp(1i*angle(source));                    % Optic or input plane (i)

    B = Rayleigh-Sommerfeld-Propagator(A,L,lambda,z);      % Propagation to the focal
                                                           plane (ii)

    C1 = (alpha.*abs(target)-beta.*abs(B)).*exp(1i.*angle(target));   % C1 forces the amplitude
                                                                      and the phase of the target in
                                                                      the signal domain.

    C2 = (gamma.*abs(B)).*exp(1i.*angle(B));               % C2 calculates the freedom
                                                           domain phase distribution.

    C = C1+C2;                                             % Signal and freedom
                                                           domains are combined.

    D = Rayleigh-Sommerfeld-Propagator(C,L,lambda,-z);     % Back-propagation from
                                                           focal plane (ii) to the optic
                                                           plane (i)

    source = D;                                            % Re-initialize for the next
                                                           loop

end
```

Algorithm S1 | The pseudo-algorithm used for synthesizing the computer generated holograms in this study.

The **alpha**, **beta**, and **gamma** are multipliers that range from 0 to 1. **Support** is a mask in the optic plane that defines the aperture. **Target** is the desired structure illumination, which is a complex field that defines the amplitude and the phase of the projected field. **Iterations** is the maximum number of iterations. **L** is the simulation size, lambda is the wavelength, **z** is the propagation distance which equals to **f**.

**abs** and **angle** are functions that convert a complex valued matrix to its amplitude and phase, respectively.

**Estimation of the diffraction efficiency**

The focal plane intensity distribution is calculated using actual material parameters first by propagating to the OSA plane which cuts off the spurious diffraction orders, and then propagating to the focal plane. Then, virtual apertures are placed around the sample and the reference beams in close proximity. This mask looks a like the target structured illumination with an enlarged for the aperture for the reference to include first few zero crossings around the sharp spot. The ratio of the integrated intensity within these regions to the total intensity that falls on to the optic multiplied by a 100 gives the focusing efficiency. Also, the ratio of the intensity within the virtual masks to that of the total intensity contained within the *P* is a measure of how good the optic reconstructs the desired intensity profile.

**Error progression as a function of the number of iterations**

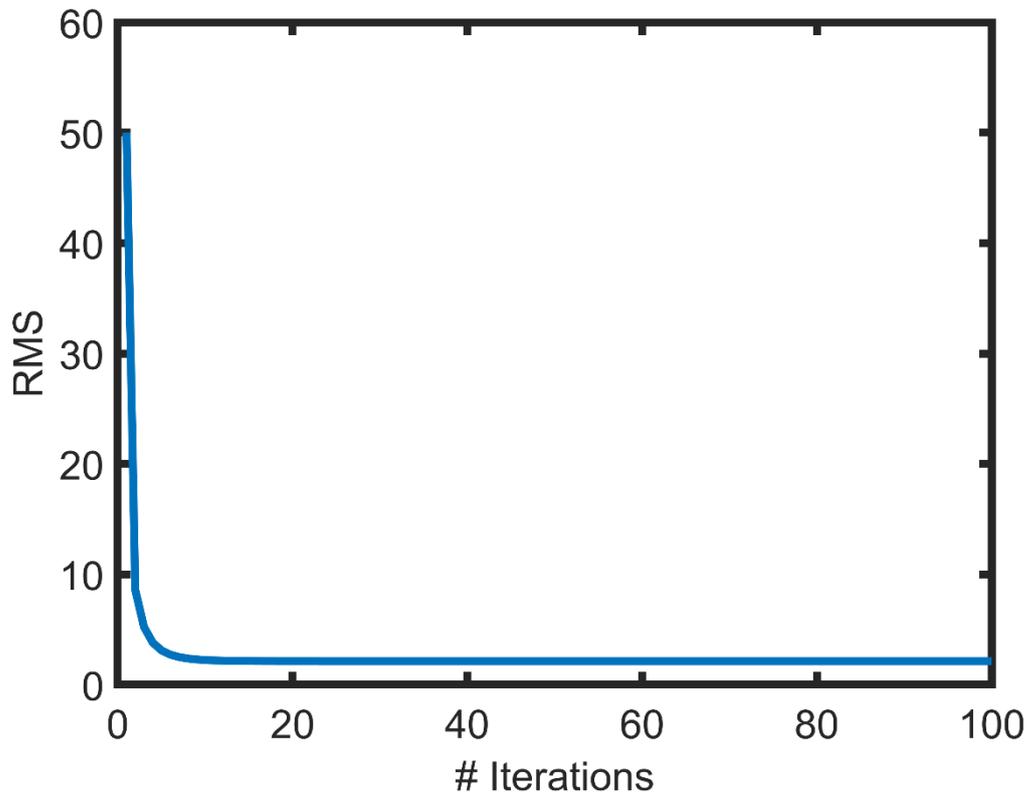

**Figure S2 | Progression of error as a function of iteration number up to 100 iterations.** Drop in RMS as iteration number increases for the following simulation parameters $L$= 200 microns, $N$= 2000, $dx$= 100 nm, $f$= 1.2 cm, CGH size= 100 microns, $\lambda$= 1 nm, $I_r$= 150, square aperture, Rayleigh-Sommerfeld propagation. Hologram synthesis with DSGSA, 20 iterations, $\alpha$= 0.5, $\beta$= 0.01, $\gamma$= 0.1.

**The effect of shot noise on reconstruction quality**

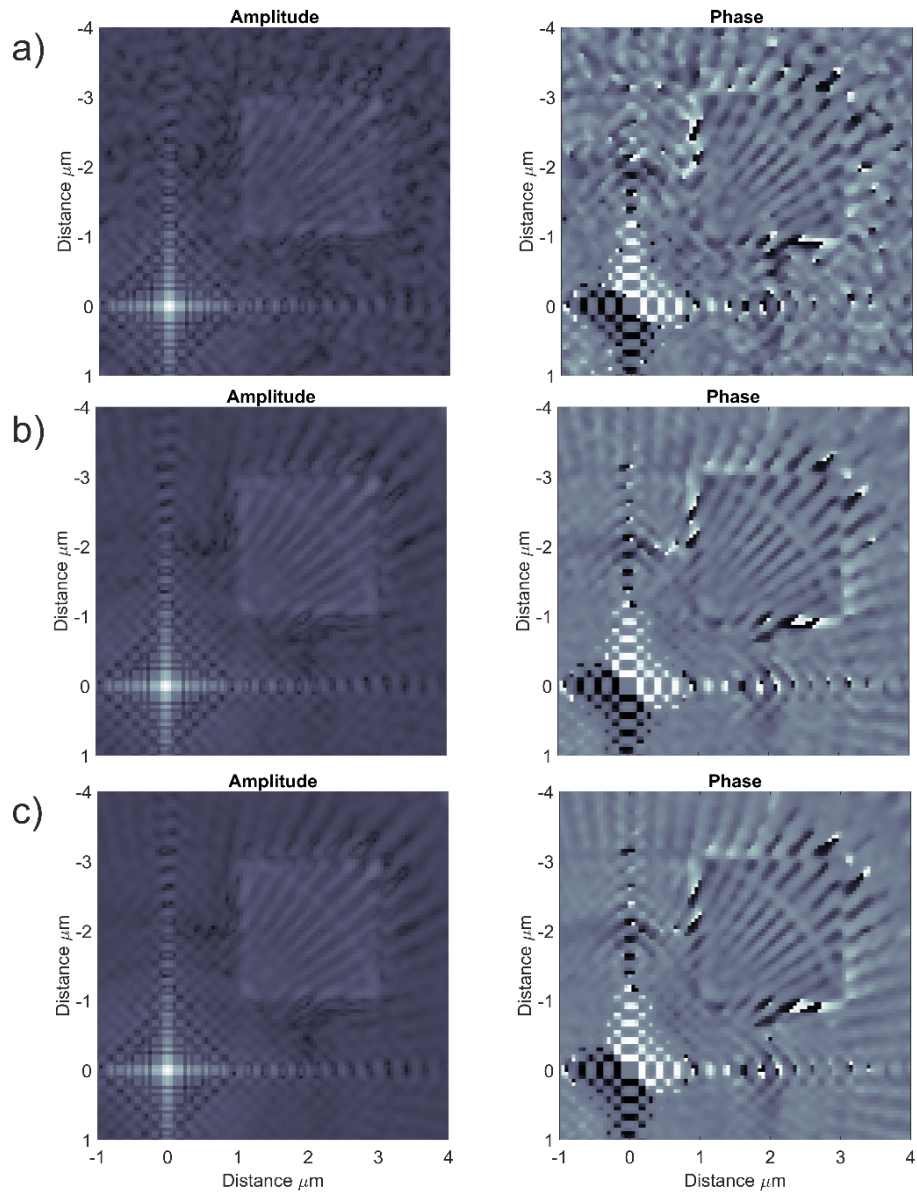

**Figure S3 | Effect of shot noise on the reconstruction quality.** Reconstructions of a portion of Siemens start from simulated data with detector Poisson noise with maximum photons of $10^1$, $10^2$ and $10^3$ per pixel. Simulations were made by Fourier propagating the exit wave to a detector 16 cm away from the sample.

**Preparation of the synthetic magnetic sample for *in silico* experiments**

X-ray scattering form factors can be expressed in the most general form as follows $f \approx f_0 + f_c + f_m$ [8], where $f_0$ is the Thomson scattering term, and $f_c$ and $f_m$ are the charge and magnetic scattering factors, respectively. The anomalous charge and magnetic components of the scattering factor $f^n(\hbar\omega)$ near an absorption edge are dependent on the incident and scattered light polarization, as discussed extensively in the literature [8, 9]. The complex refractive index including the magnetic contributions is then $\tilde{n}(\hbar\omega) = 1 - n_a r_e \lambda^2 / 2\pi \, [f^n(\hbar\omega)]$, where $n_a$ is the number density of atoms in the solid and $r_e$ is the classical electron radius.

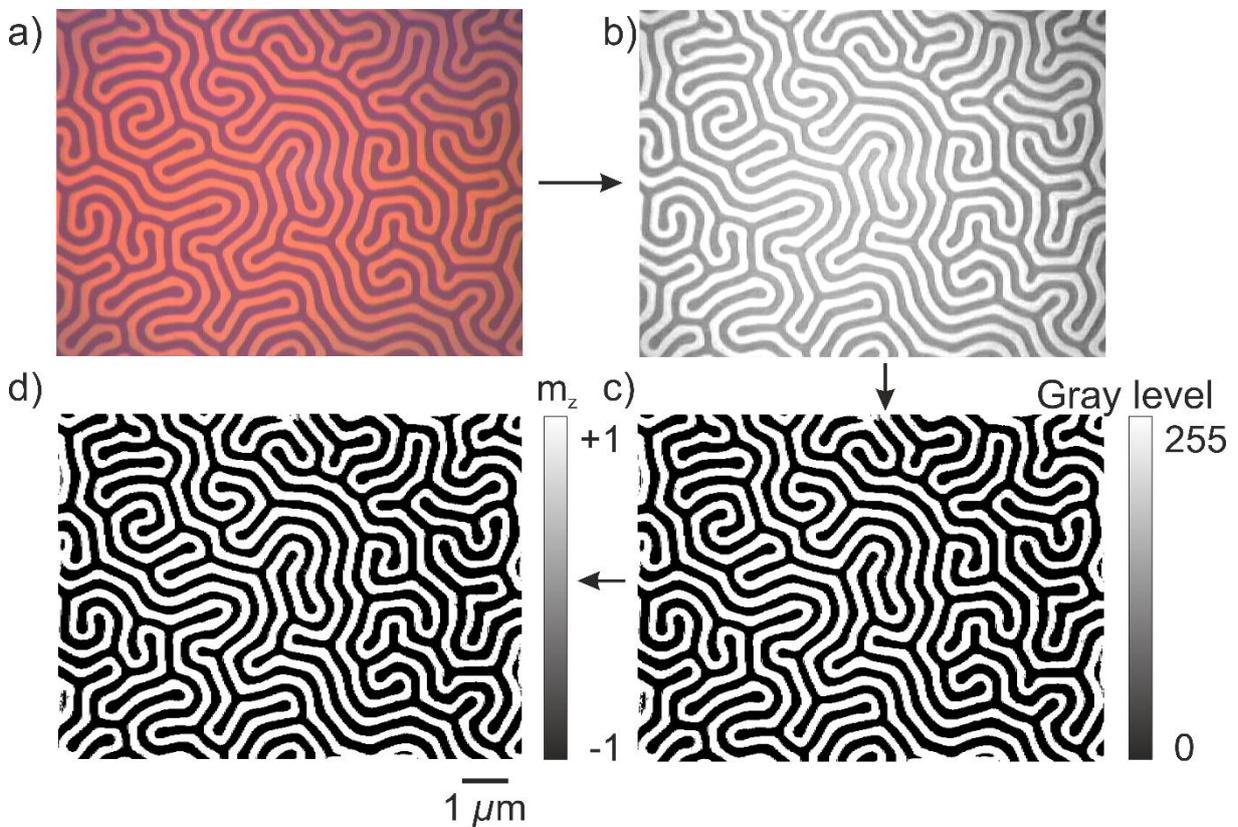

**Figure S4 | Image processing history of worm-domain image. a** A magneto-optic image of a bismuth-iron-garnet thin film (see below explanation). The image was used as a template for simulating the magnetization directions. Original image was 640 × 480 pixels and the domains were reported to be 3-μm wide. **b, c** and **d** are 8-bit, binarized and re-scaled versions, respectively. Image in **d** corresponds to the $m_z$.

Worm-domain type magnetization shown in Figure 3a in the main text was formed by starting from an magneto-optic (Faraday rotation) image (Figure S5) of an epitaxial bismuth-iron-garnet thin film layer (image source: https://commons.wikimedia.org/wiki/File:Magnetic_stripe_domains.jpg, final access date: 10.02.2020 DMGualtieri [CC BY-SA (https://creativecommons.org/licenses/by-sa/3.0)]. This image was converted to 8-bit and binarized by applying a threshold in ImageJ. Then, the resulting binary image was imported to MATLAB® and scaled between -1 and +1 representing an $m_z$ that goes into and out of the paper plane (designated as out-of-plane magnetization). The resulting array representing the $m_z$ was resampled to have the same step-size as the simulation domain. Finally, the complex refractive index of the sample was calculated by using the Fe molar density, charge and magnetic scattering factors taken from Daan B. Boltje's master's thesis [10].

**Effect of the beamstop on the structured illumination**

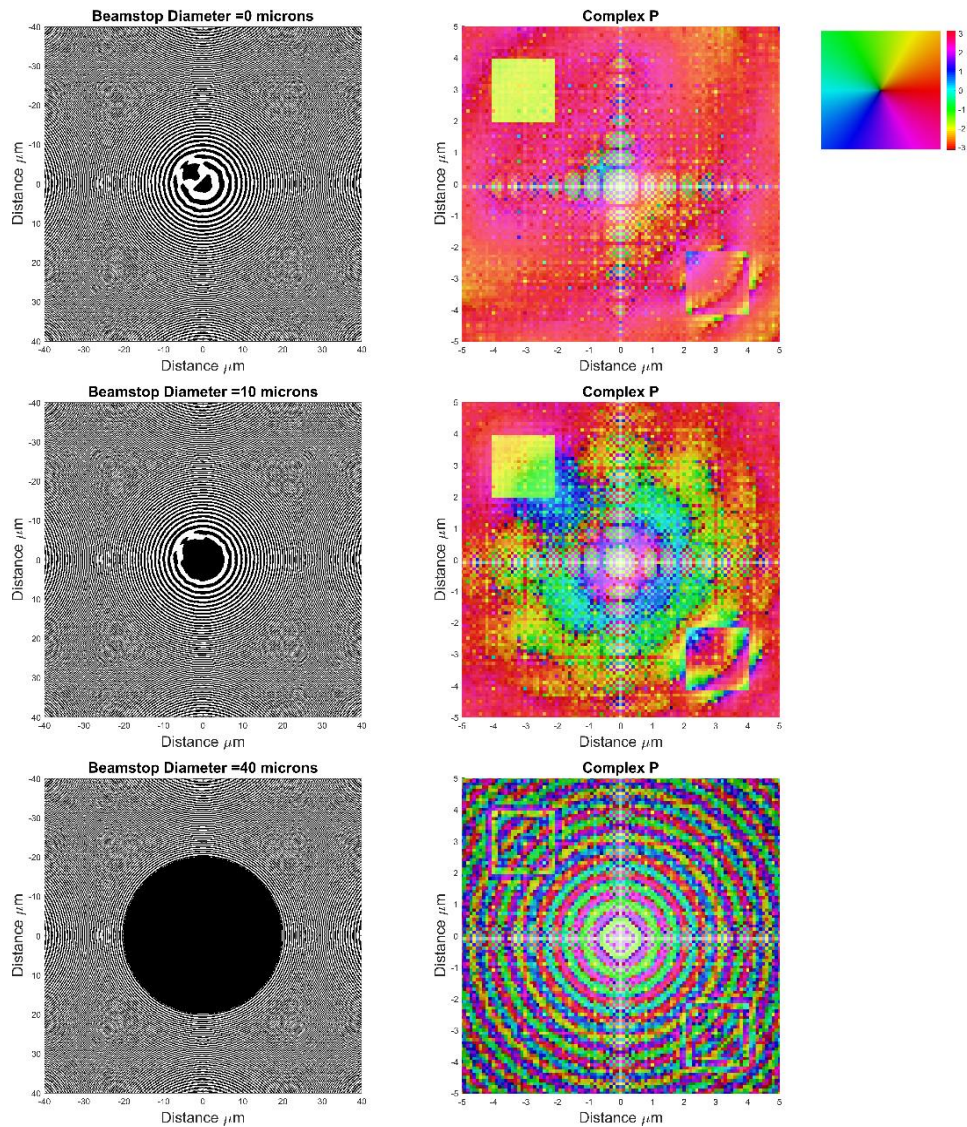

**Figure S5 | Effect of introduction of a beamstop on the complex structured illumination function at the focal plane.** Left column top to bottom shows the transmission function of the CGH with increasing beamstop diameter. Right column top to bottom shows the domain-coloring plot of the complex plane at focus corresponding to the CGH on the left. Inset shows the color that corresponds to phase. Brightness in the images correspond to the intensity value. Simulation parameters $L$= 200 microns, $N$= 2000, $dx$= 100 nm, $f$= 1.2 cm, CGH size= 100 microns, $\lambda$= 1 nm, $I_r$= 150, square aperture, Rayleigh-Sommerfeld propagation. Hologram synthesis with DSGSA, 20 iterations, $\alpha$= 0.5, $\beta$= 0.01, $\gamma$= 0.1.

**References to the SI**

1. D. G. Voelz, *Computational fourier optics: a MATLAB tutorial* (SPIE, 2011).
2. K. Li, M. Wojcik, and C. Jacobsen, "Multislice does it all - calculating the performance of nanofocusing X-ray optics," Opt. Express **25**, 1831-1846 (2017).
3. C. Chang, J. Xia, L. Yang, W. Lei, Z. Yang, and J. Chen, "Speckle-suppressed phase-only holographic three-dimensional display based on double-constraint Gerchberg-Saxton algorithm," Appl. Opt. **54**, 6994-7001 (2015).
4. K. Keskinbora, C. Grévent, U. Eigenthaler, M. Weigand, and G. Schütz, "Rapid Prototyping of Fresnel Zone Plates via Direct Ga+ Ion Beam Lithography for High Resolution X-Ray Imaging," ACS Nano **7**, 9788-9797 (2013).
5. K. Keskinbora, U. T. Sanli, M. Baluktsian, C. Grévent, M. Weigand, and G. Schütz, "High-throughput synthesis of modified Fresnel zone plate arrays via ion beam lithography," Beilstein Journal of Nanotechnology **9**, 2049-2056 (2018).
6. L. Loetgering, M. Baluktsian, K. Keskinbora, R. Horstmeyer, T. Wilhein, G. Schütz, K. S. E. Eikema, and S. Witte, "Generation and characterization of focused helical x-ray beams," Science Advances **6**, eaax8836 (2020).
7. A. L. Levitan, K. Keskinbora, U. T. Sanli, M. Weigand, and R. Comin, "Single-frame far-field diffractive imaging with randomized illumination," Opt. Express **28**, 37103-37117 (2020).
8. J. P. Hannon, G. T. Trammell, M. Blume, and D. Gibbs, "X-Ray Resonance Exchange Scattering," Phys. Rev. Lett. **61**, 1245-1248 (1988).
9. S. Eisebitt, M. Lörgen, W. Eberhardt, J. Lüning, J. Stöhr, C. T. Rettner, O. Hellwig, E. E. Fullerton, and G. Denbeaux, "Polarization effects in coherent scattering from magnetic specimen: Implications for x-ray holography, lensless imaging, and correlation spectroscopy," Phys. Rev. B **68**, 104419 (2003).
10. D. B. Boltje, *Voltage Induced Near Interface Changes of the Magnetocrystalline Anisotropy Energy: A study by X-ray Resonant Techniques in Combination with Conventional Magnetometry* (Universität Stuttgart, 2017).